\documentclass[preprint,12pt]{elsarticle}
\usepackage{graphicx}
\usepackage{color}
\usepackage{amsmath}
\usepackage{amssymb}
\usepackage{natmove}
\usepackage{natbib}
\usepackage{hyperref}
\usepackage{bm}
\usepackage{soul}
\usepackage{comment}

\journal{Materials Today Physics}

\begin{document}

\begin{frontmatter}

\title{Structural and electronic transformations in TiO$_2$ induced by electric current}

\author[1]{Tyler C. Sterling\corref{cor}}
\ead{ty.sterling@colorado.edu}
\author[2]{Feng Ye}
\author[3]{Seohyeon Jo}
\author[1]{Anish Parulekar}
\author[1]{Yu Zhang}
\author[1,4]{Gang Cao}
\author[5]{Rishi Raj}
\author[1,4]{Dmitry Reznik\corref{cor}}
\ead{dmitry.reznik@colorado.edu}
\cortext[cor]{Corresponding Author}

\affiliation[1]{organization={Department of Physics, University of Colorado Boulder}, city={Boulder}, postcode={80309}, state={Colorado}, country={USA}}
\affiliation[2]{organization={Neutron Scattering Division, Oak Ridge National Lab}, city{Oak Ridge}, postcode={37830}, state={Tennessee}, country={USA}}
\affiliation[3]{organization={Materials Science and Engineering, University of Colorado Boulder}, city={Boulder}, postcode={80309}, state={Colorado}, country={USA}}
\affiliation[4]{organization={Center for Experiments on Quantum Materials, University of Colorado Boulder}, city={Boulder}, postcode={80309}, state={Colorado}, country={USA}}
\affiliation[5]{organization={Department of Mechanical Engineering, University of Colorado Boulder}, city={Boulder}, postcode={80309}, state={Colorado}, country={USA}}

\date{\today}

\begin{abstract}
In-situ diffuse neutron scattering experiments revealed that when electric current is passed through single crystals of rutile TiO$_2$ under conditions conducive to flash sintering, it induces the formation of parallel planes of oxygen vacancies. Specifically, a current perpendicular to the $c$-axis generates planes normal to the (132) reciprocal lattice vector, whereas currents aligned with the $c$-axis form planes normal to the (132) and to the (225) vector. The concentration of defects increases with incresing current. The structural modifications are linked to the appearance of signatures of interacting Ti$^{3+}$ moments in magnetic susceptibility, signifying a structural collapse around the vacancy planes. Electrical conductivity measurements of the modified material reveal several electronic transitions between semiconducting states (via a metal-like intermediate state) with the smallest gap being 27 meV. Pristine TiO$_2$ can be restored by heating followed by slow cooling in air. Our work suggests a novel paradigm for achieving switching of electrical conductivity related to the flash phenomenon.
\end{abstract}

\begin{graphicalabstract}
\begin{figure*}
    \centering
    \includegraphics[width=\linewidth]{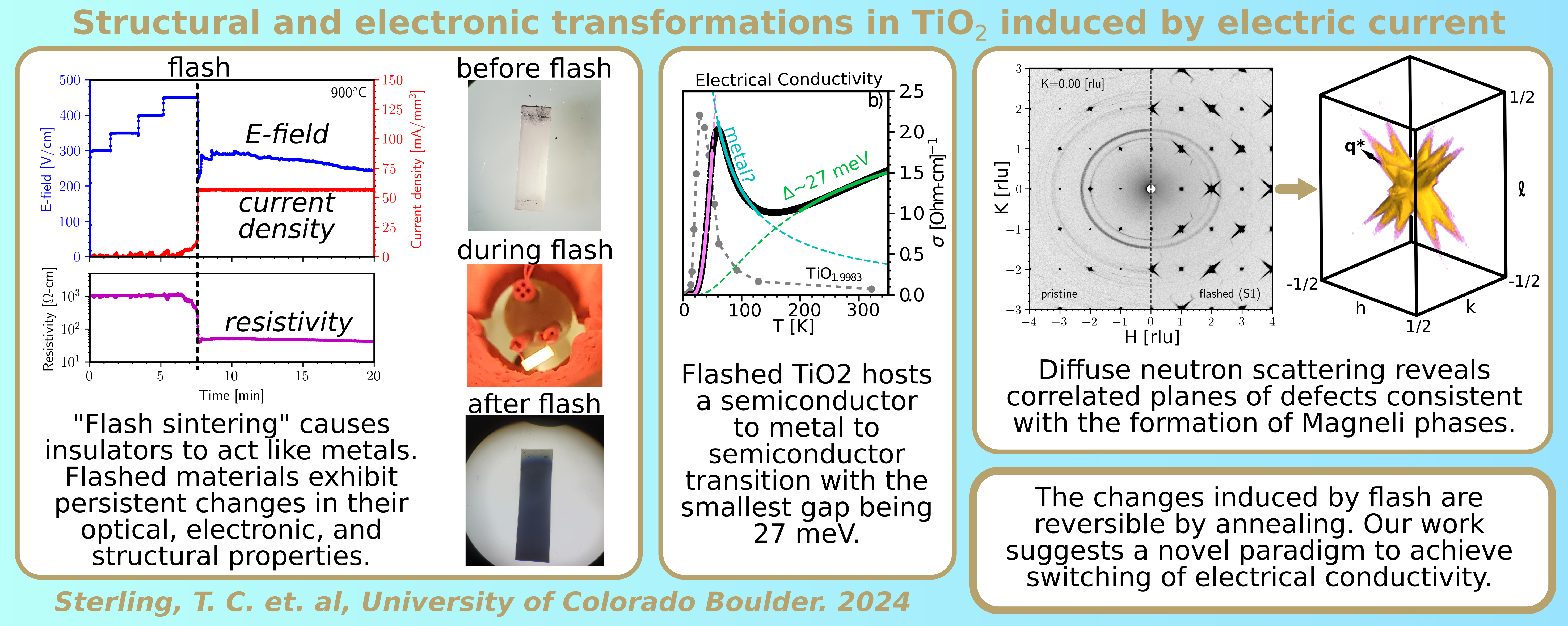}
\end{figure*}
\end{graphicalabstract}

\begin{highlights}
\item ``Flash sintering" makes insulators behave like metals
\item Flash causes persistent but reversible changes to some materials
\item Diffuse scattering shows that flash creates extended defects in rutile TiO$_2$
\item The electrical conductivity of flashed rutile is significantly modified
\end{highlights}


\end{frontmatter}

\section{Introduction}

\begin{figure*}
    \centering
    \includegraphics[width=\linewidth]{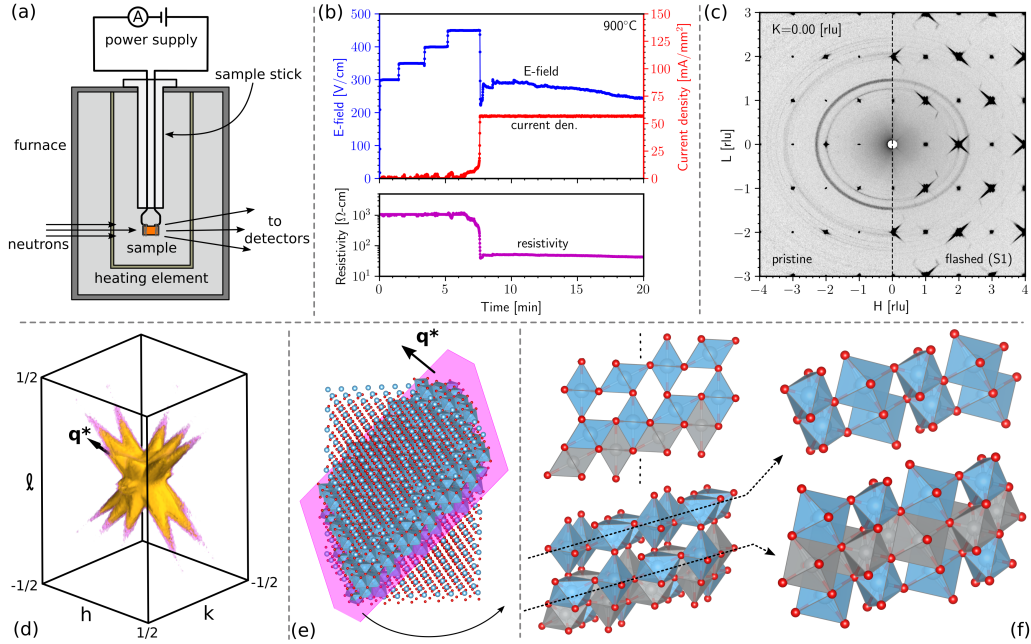}
    \caption{(a) Schematic of the in-situ neutron scattering flash experiment under constant-current conditions like in the typical I-V curve during flash in (b). When the E-field is increased to a critical value, the resistivity drops sharply ($\sim$ an order of magnitude); this occurs at about 7.5 minutes. (c) Scattering pattern from pristine (left) and flashed (right) samples in the $H$-$L$ plane with $K=0$. (d). Diffuse scattering features from the S1 dataset in (c) translated into the 1$^{st}$ Brillouin zone and summed as discussed in the text. (e) A model of a Magn\'eli phase shear plane as described in ref. \cite{bursill1971crystal}. The vector $\bm{q}^*$ in (d) and (e) points along $(1\bar{3}2)$ and the displacement vector for the shear in (e) is $1/2[011]$. (f) A model of the defect region in (e) showing different views (left) and an exploded view (right). The shear plane consists of chains of face-sharing octahedra (gray) separated by rutile-like edge sharing octahedra (blue). Between shear planes, the structure is rutile.}
    \label{fig:main_fig}
\end{figure*}

Flash sintering has emerged as a transformative technique for material processing, revolutionizing the synthesis and densification of ceramic materials \cite{yadav2023room,raj2021flash,raj2016analysis,raj2012joule,cologna2010flash}. This method involves the application of an electric field to significantly reduce the sintering time and temperature, achieving results comparable to traditional methods in mere seconds. This not only conserves energy but also paves the way for the creation of materials with novel properties.

In this study, we applied an electric field to rutile TiO$_2$ single crystals under flash sintering conditions, rapidly modifying the atomic structure and switching electrical resistivity from a high to low resistance state with novel temperature dependence. Our in-situ neutron diffuse scattering measurements reveal that this process consistently generates oriented parallel planes of structural defects, likely oxygen vacancies, which remain stable at room temperature. This altered form of TiO$_2$ shows several electronic transitions as a function of temperature and, at room temperature, exhibits a bandgap reduced from $\sim 3.05$ eV \cite{cronemeyer1952electrical,amtout1995optical} in the pristine material to 27 meV. Additionally, magnetic susceptibility measurements indicate an enhanced non-Curie-Weiss magnetic susceptibility, consistent with the presence of Ti$^{3+}$ ions with interacting moments, suggesting a structural collapse around the planes of oxygen vacancies. The formation of interacting moments by electric current suggests flash can be used to induce novel electronic physics. These findings offer a promising pathway towards engineering new devices, such as memristors, through the current-induced introduction of defects in TiO$_2$, opening new avenues for advanced material applications.

\section{Methods}

To understand the structural modifications of materials during flash, we designed and built an apparatus for in-situ neutron scattering measurements under controlled E-field and high temperatures at the CORELLI diffractometer at the Spallation Neutron Source (SNS) at Oak Ridge National Laboratory. This instrument combines the high efficiency of white beam Laue diffraction with energy discrimination, allowing us to measure both elastic and total scattering in a single experiment over large volumes of reciprocal space. It also accesses large momentum transfer Q and high momentum resolution to distinguish diffuse signals from nearby Bragg peaks \cite{ye2018implementation}.

Elastic neutron scattering experiments were performed on rutile TiO$_2$ single crystals with current applied along the $b$- and $c$-axes. For current along the $b$-axis, we measured a crystal in-situ and a pre-flashed one ex-situ, which was quenched in LN2 after being flashed in air. The in-situ experiments were conducted at multiple temperatures and current densities to study the dependence of scattering on flash conditions. For current along the $c$-axis, we measured one sample and one current density, with additional measurements on the $c$-axis flashed sample ex-situ using a low-background close cycle refrigerator to characterize structural modifications. In this study, we focus primarily on two samples: sample-1 (S1), flashed in \emph{air} with current along the $b$-axis and then quenched, and sample-2 (S2), flashed in \emph{vacuum} at the SNS with current along the $c$-axis. The dimensions of all samples were $10\times3\times 1$ mm.

Refining the atomic coordinates via analysis of Bragg intensities results in significant uncertainty due to experimental conditions being optimized for diffuse scattering measurements. We will address structural refinement following planned additional measurements, which are beyond the scope of the current project.

Temperature dependence of electrical resistivity and magnetic susceptibility was measured on a pristine single crystal and on S2 using the Quantum Design MPMS system. S1 broke during the neutron scattering experiments, rendering the pieces too small for reliable transport measurements. The conductivity of the pristine sample was too low to measure, and the flashed sample was too thin to measure conductivity in the $ab$ plane. The sample environment at CORELLI does not support logging I-V data, so we performed an offline experiment to measure the I-V curve in fig. \ref{fig:main_fig}b on a different sample under similar flash conditions to emphasize the non-linear change in resistivity that is typical during flash experiments \cite{yadav2023room,raj2016analysis}.

\section{Results}

During flash sintering \cite{raj2016analysis}, a sample is first held at an elevated temperature, and an electric field (E-field) is applied across it (fig.  \ref{fig:main_fig}a). The E-field is increased continuously or stepwise until the resistivity begins to fall nonlinearly, which in fig.  \ref{fig:main_fig}b occurs at 6 minutes when the E-field reaches 150 V/cm. Eventually, at around 8 minutes, the sample abruptly switches to a low-resistance state, and the power supply reaches the current limit. At this point, the voltage is automatically regulated to maintain constant current, signaling the onset of flash. Alternatively, a constant E-field can be applied, and the temperature increased to a critical value where the resistivity drop occurs. In the flash state, the sample usually glows brightly due to electroluminescence. This process exhibits some material-dependent variability; in some cases, flash occurs under moderate critical E-field strengths at room temperature. Furthermore, this glow might not always be visible, depending on the luminescence spectrum and brightness, which vary with material properties \cite{roy2023structural}.

Our experiments show that TiO$_2$ loses some oxygen during flash. Turning off the current and slowly cooling the sample in air often leads to re-absorption of oxygen, restoring the pristine high-resistance state. Remarkably, the sample remains in the flash state under steady-state current conditions, even if removed from the furnace. Immersing the flashing sample in liquid nitrogen (LN2) followed by shutting off the current, or conducting flash sintering in a vacuum, prevents oxygen re-absorption.

\begin{figure}
    \centering
    \includegraphics[width=0.65\linewidth]{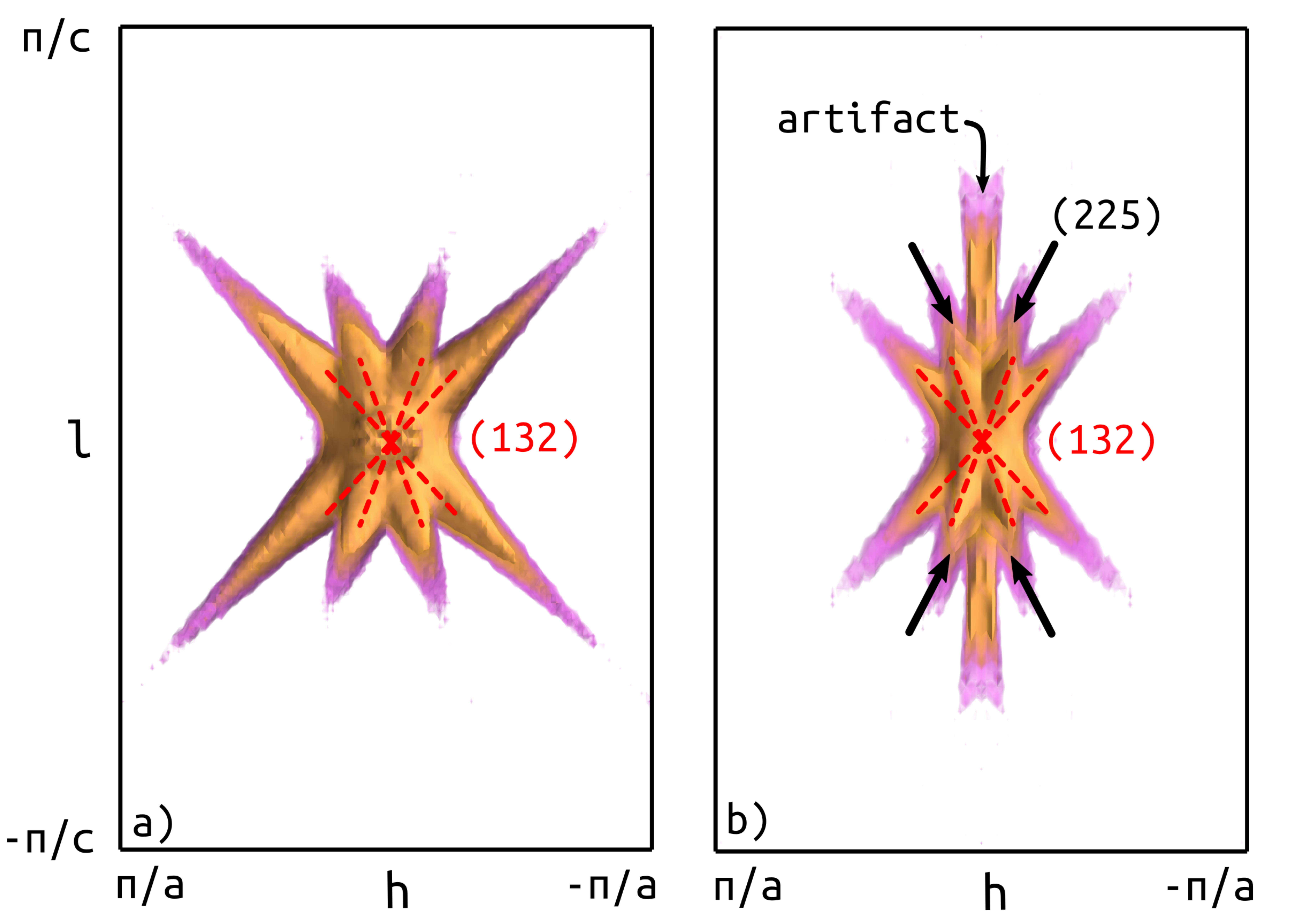}
    \caption{Volumetric plots of the summed S1 (a) and S2 (b) datasets. The dashed red lines label the (132) features, whereas black arrows label the (225) features. (a) The sample was flashed and quenched in LN2 and the current was perpendicular to the $b$-axis. (b) The sample was flashed in vacuum and then was allowed to cool. Current was parallel to the $c$-axis. The artifact labeled in (b) is from neutrons saturating the detectors; it is not physically interesting.}
    \label{fig:summed_views}
\end{figure}

Our neutron scattering experiments provide significant insights into the structural transformations in TiO$_2$ under the influence of electrical current, with profound implications for understanding flash-induced modifications. We observed that Bragg peak positions consistently aligned with the tetragonal symmetry of the rutile structure across all datasets, indicating no symmetry breaking of the average structure upon the application of current. The primary changes induced by the Flash technique were confined to the local structure.

In this study, we focus primarily on two samples: sample-1 (S1), flashed in \emph{air} with current along the $b$-axis and then quenched, and sample-2 (S2), flashed in \emph{vacuum} at the SNS with current along the $c$-axis. We also discuss several in-situ measurements with current along the $b$-axis (see methods for details). The results of in-situ neutron scattering measurements with electrical current passing through the samples were qualitatively similar to those obtained from ex-situ flashed samples, with a notable difference for sample S2. Detailed results for the in-situ measurements at several temperatures and currents are provided in the supplementary information \cite{supp_info}. In the main text, we mainly focus on ex-situ measurements of samples S1 and S2 as quantitative analysis of the in-situ data was not possible due to the large furnace background.

As shown in fig.  \ref{fig:main_fig}, the primary effect of the flash process is the emergence of needle-like shoulders on the Bragg peaks, extending in all symmetry-equivalent (132) directions. Both S1 and S2 samples exhibited these (132) features; however, S2 also displayed less pronounced needles oriented along (225), as illustrated in fig. \ref{fig:summed_views}b. The differences in the apparent widths of features in fig.  \ref{fig:summed_views}a versus fig.  \ref{fig:summed_views}b are attributed to instrument resolution, as the two samples were mounted with different orientations in the scattering plane. In what follows, we use Miller indices in the basis of primitive rutile lattice vectors (tetragonal symmetry with $a=b\neq c$).

\begin{figure}
    \centering
    \includegraphics[width=0.65\linewidth]{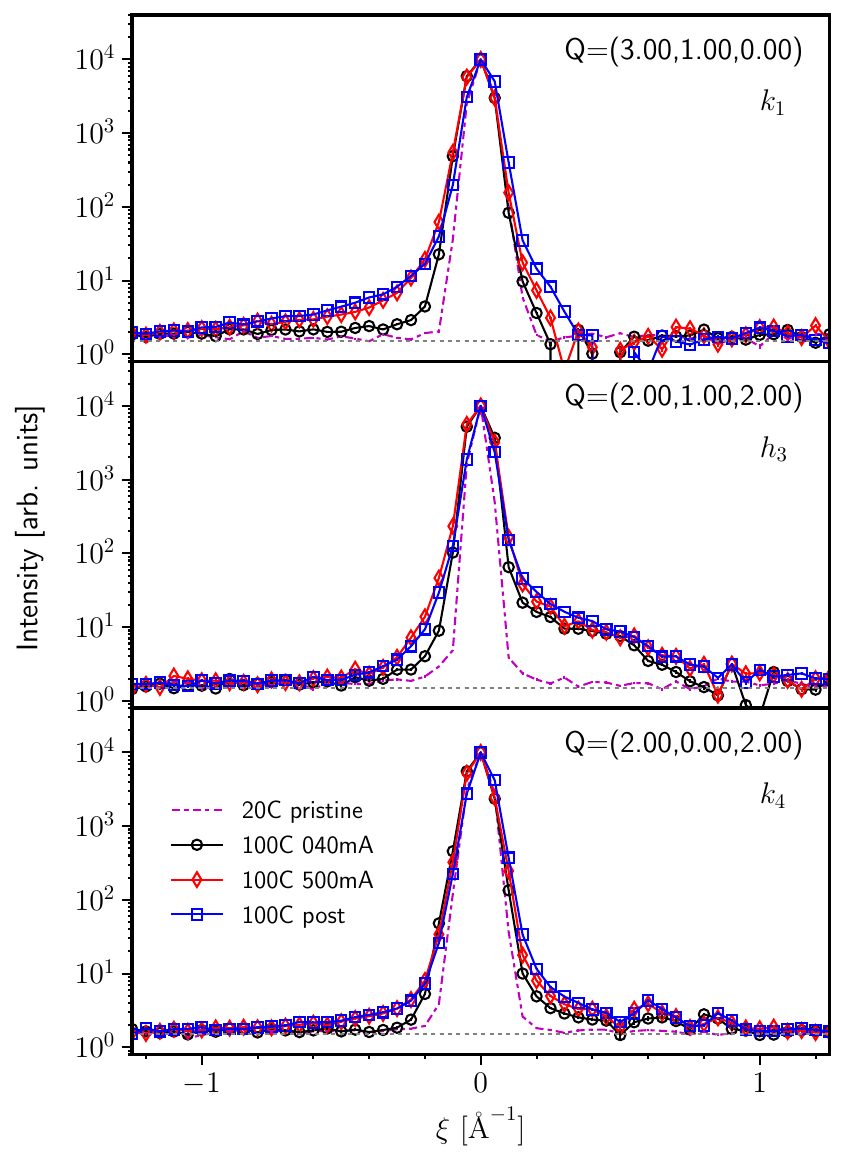}
    \caption{Current and temperature dependence of the diffuse scattering features along different needles thru different Bragg peaks labeled in the figure. Background was determined by taking cuts through nearby regions with no diffuse scattering and then subtracted from the data.  Sample ``$100^\circ$C post" is the in-situ flashed crystal after the field was turned off and the sample cooled (under vacuum) back to the furnace temperature $100^\circ$C. Background due to the furnace and electrodes was subtracted. Since the needles aren't oriented along a high-symmetry direction, we plot data in $\textrm{\AA}^{-1}$. See the text and supp. info \cite{supp_info} for detailed explanation of the coordinates and orientation of the needles.}
    \label{fig:combined_line_cuts}
\end{figure}

The intensity and asymmetry of the needles varied from Bragg peak to Bragg peak due to structure factor effects. By projecting all diffuse scattering features into the first Brillouin zone, a sea urchin-like pattern emerged. We picked all Brillouin zones in each data set with significant diffuse scattering, translated these scattering intensities to the 1$^{st}$ Brillouin zone, then summed them (see fig.  \ref{fig:summed_views} and supp. info \cite{supp_info}). We call different symmetry-equivalent needles $k_{1}$, $k_2$, $...$, $h_1$, $...$, etc \cite{supp_info}. The sea-urchin structure is present for all currents and temperatures that we measured. The directions of the needles were determined by fitting the coordinates of their intersections with integrated planes of intensity cuts through the summed data. These needle directions are all equivalent under tetragonal symmetry operations. The direction of the needles is independent of current (see supp. info. \cite{supp_info}).

In fig. \ref{fig:combined_line_cuts}, we present current and temperature-dependent cuts along several different needles. These cuts are made along straight lines in reciprocal space that align with the needle orientations, rather than high-symmetry directions, and are presented in $\textrm{\AA}^{-1}$. Detailed fitting of the needle directions and additional cuts along needles are provided in the supplementary information \cite{supp_info}.

It is well-established that needle-like structures in diffraction correspond to planar correlations in real space. The primary effect of flash on TiO$_2$ is to produce either pancake-like clusters of defects and/or planar slip dislocations oriented perpendicular to the needles. Analysis of the S1 dataset, where we have the most comprehensive data, allowed us to estimate the correlation lengths of the defects to be approximately $46~\textrm{\AA}$ perpendicular to the pancake planes and $\sim200~\textrm{\AA}$ within the planes (line cuts perpendicular to the needles are resolution-limited: see supp. info \cite{supp_info}).

Figure \ref{fig:combined_line_cuts} compares the scattering intensity distribution for representative cuts along the needles of the in-situ sample at different electrical current densities and temperatures in the furnace. The pristine sample exhibited only resolution-limited Bragg peaks, with no diffuse scattering. In contrast, flashed samples displayed diffuse scattering tails on the Bragg peaks. Although needle orientation was independent of current (see supp. info. \cite{supp_info}), the diffuse intensity increased with higher current density. For identical sample volumes, which is the case for our experiments, an increase in diffuse scattering intensity corresponds to an increase in defect concentration. This increase was not linear; for example, 40 mA at 100$^\circ$C produced less diffuse intensity than higher currents, where the effect seemed to saturate. During the flash process, the sample temperature rises considerably due to heating by electric current and experiments to accurately characterize this heating are planned. Preliminary estimates based on Bragg peak intensities show that 500 mA results in extra heating of several hundred degrees ($^\circ$C). Despite this, the diffuse scattering pattern remains consistent with that of the post-flashed sample, which experienced no additional heating. This suggests that temperature does not significantly influence the diffuse scattering pattern.

\begin{figure}
    \centering
    \includegraphics[width=0.65\linewidth]{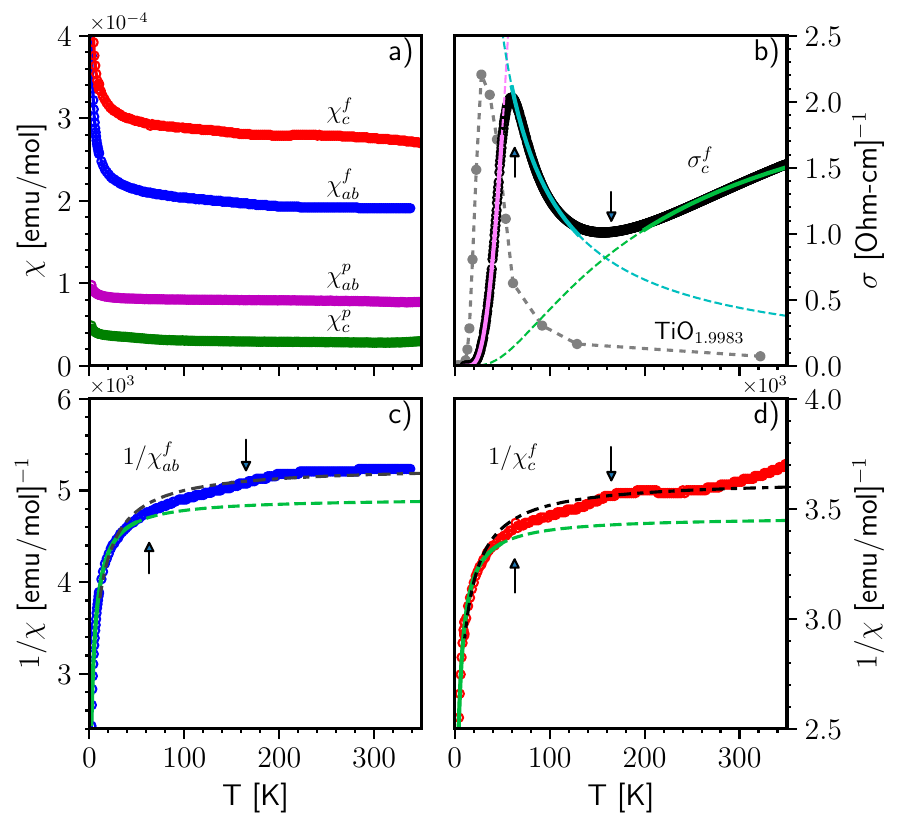}
    \caption{Magnetic susceptibility $\chi$ (a), electrical conductivity $\sigma$ (b), and $1/\chi$ (c-d) of pristine and flashed S2 sample (see text). $\chi^f_{ab}$ and $\chi^f_c$ are the susceptibilities of the flashed sample in the $ab$ plane and along the $c$-axis respectively. $\sigma^f_{c}$ is the conductivity along the $c$-axis of the flashed sample. The arrow indicates the position of the `dip' in the conductivity which coincides with anomalies in the susceptibility. The different curves plotted with $\sigma$ and $1/\chi$ corresponds to fits through different temperature regimes as discussed in the text. The gray markers in b) are conductivity data for TiO$_{1.9983}$ from ref. \cite{hasiguti1994electrical} divided by 10.}
    \label{fig:mat_props}
\end{figure}

The temperature dependence of the electrical conductivity of S2 (fig. \ref{fig:mat_props}b) at low temperatures $T \lesssim$ 50 K fits the semiconductor model $\sigma(T) = \sigma_0 \exp(-E_A/k_B T) $ with $E_A = $ 13.5 meV and $\sigma_0 = $ 39.7 (Ohm-cm)$^{-1}$. At high temperatures $T \gtrsim$ 150 K, $E_A = $ 15.2 meV and $\sigma_0 = $ 2.5 (Ohm-cm)$^{-1}$, although the fit is less precise. We estimate the band gap, $\Delta$, from $E_A = \Delta/2$. An alternative approach might involve assuming multiple gaps in the high-temperature regime \cite{yagi1996electronic,hasiguti1994electrical}.

In the intermediate temperature range (60 K $\lesssim T$ $\lesssim$ 150 K), the conductivity fits the model $\sigma(T) = 1/ [\rho_D + \rho_0 (T/\Theta)^n]$, consistent with metallic behavior. Here, $\rho_D$ represents the resistivity due to defect scattering, $\rho_0$ is the characteristic resistivity from other scattering forms, $\Theta$ is the temperature scale, and $n$ is the exponent. At low temperatures, $n = 5$ corresponds to electron-phonon scattering and $n = 2$ to electron-electron scattering, though $n = 1$ is often used empirically for small temperature changes. We fit $\rho_D,~\rho_0,$ and $n$. We find $\rho_D \approx$ 0 (Ohm-cm) an $n\approx 1$, so that $\sigma(T)\sim \sigma_0 \times (\Theta/T)$ with $\sigma_0 = 1/\rho_0 = $ 0.2 (Ohm-cm)$^{-1}$ and $\Theta =$ 654.5 K. Interestingly, $\Theta$ is similar to but lower than the known Debye temperature of pristine rutile TiO$_2$, $\Theta_D=$ 943 K \cite{wu1982elastic}, though the interpretation with $n\equiv 1$ is not physically meaningful. 

The magnetic susceptibilities were analyzed assuming Curie-Weiss behavior to extract magnetic moments ($\mu_{eff} = \sqrt{8 C}$) and Curie-Weiss temperatures ($\Theta_{CW}$). A temperature-independent contribution ($\chi_0$) was also included in the susceptibility model: $\chi(T) = \chi_0 + C/(T-\Theta_{CW})$. Above approximately 60 K, the susceptibilities of the flashed sample deviated significantly from Curie-Weiss behavior (Figs. \ref{fig:mat_props}c-d). We separately fit both the Curie-Weiss-like low-temperature regimes (dashed lines in Figs. \ref{fig:mat_props}c and d) and the full curves (dashed-dotted lines in Figs. \ref{fig:mat_props}c-d). Only the parameters extracted from the low-T regime for the flashed samples were physically sensible, and we focus on these here. Fits to the full temperature range are provided in the supplementary information\cite{supp_info}. As a reference, the pristine samples followed Curie-Weiss behavior up to $T\sim$300 K.

At low temperature for the flashed sample, $\mu_{eff} = 0.069~\mu_B$/Ti in the $ab$ plane and $\mu_{eff} = 0.065~\mu_B$/Ti along the $c$-axis for the flashed sample. The increase in the effective moment in S2 is consistent with the emergence of Ti$^{3+}$(3d$^1$) ions; the occurrence of Ti$^{3+}$ is typical of reduced rutile TiO$_2$ as shown before with x-ray photoelectron spectroscopy results from both electroformed TiO$_2$ \cite{szot2011tio2} and bulk reduced rutile Ti$_4$O$_7$ Magn\'eli phases \cite{kim2023performance,taguchi2010anomalous}. $\chi_0$ and $\Theta_{cw}$ are provided in the supp. info. \cite{supp_info}. In the pristine sample $\mu_{eff} = 0.024~\mu_B$/Ti in the $ab$ plane and $\mu_{eff} = 0.025~\mu_B$/Ti along the $c$-axis.

Both susceptibilities of the flashed sample show anomalies at $T \approx 165$ K (down-arrows in figs. \ref{fig:mat_props}b-d). Along the $c$-axis, there is a dip in the susceptibility; in the $ab$ plane, there is a ``kink". There are no corresponding anomalies in the susceptibilities of the pristine sample. Intriguingly, a minimum in the conductivity appears at the same temperature (\ref{fig:mat_props}b). There is also a maximum in the conductivity at $T\approx 60$ K (up-arrow in fig.  \ref{fig:mat_props}b); $T\approx 60$ K is where the susceptibilities of the flashed sample begin to deviate significantly from Curie-Weiss behavior (up-arrows in Figs. \ref{fig:mat_props}c-d). These findings reveal complex and nuanced changes in TiO$_2$ as a result of flash.

\section{Discussion and Conclusions}

Flash/quench treatment induces radical yet reproducible and potentially controlled modifications in atomic structure and bulk properties. In our study, we demonstrated that this treatment introduces planar defects into TiO$_2$, transitioning it to a low-resistance phase. Insulators, such as yttria-stabilized zirconia, often become metallic under similar conditions. In the quasi-2D quantum material Pr$_2$CuO$_4$, this treatment enhances oxygen vacancy ordering, thereby increasing the three-dimensionality of electrical resistivity and magnetization, and sharpening of Raman-active phonons \cite{roy2023structural}.

Despite the enigmatic nature of the atomistic mechanism behind flash, it has been successfully applied across a broad spectrum of materials, from metals like aluminum \cite{mcwilliams2018sintering,raj_private} and tungsten \cite{raj_private} to both simple and complex oxides \cite{raj2021flash}. Typically, ceramics requiring hours at approximately $\sim$1500$^\circ$C  to sinter can now be consolidated from powders in seconds at significantly lower temperatures under flash conditions. Recent advancements in flash sintering include synthesizing new materials, such as high-entropy oxides with exceptional lithium-ion conductivity for next-generation batteries \cite{yan2022reactive}, and transforming insulators into metals. Initially observed in polycrystalline materials and attributed to Joule heating at grain boundaries \cite{yu2017review}, flash sintering's discovery in single crystals \cite{yadav2017onset} revealed it as a bulk effect.

Rutile TiO$_2$, a wide-gap insulator transparent to visible light, becomes electrically conducting and light-absorbing when slightly reduced to TiO$_{2-x}$ ($x<2$) \cite{bursill1969new,hasiguti1970electrical,yagi1996electronic}.  This transformation is due to the formation of Magn\'eli phases, where oxygen vacancies aggregate into semiconducting planes, causing shear dislocation of adjacent rutile slabs \cite{bursill1969new, bursill1971aggregation, bursill1971crystal, bursill1970displacement, zhang2021andersson, liborio2008thermodynamics}. The chemical formulae of the Magn\'eli phases are Ti$_n$O$_{2n-1}$, with the orientation and density of defect planes depending on $n$. For commonly studied more-reduced phases ($4 \leq n \leq 10$), the shear planes are perpendicular to $(121)$ \cite{andersson1960crystal,zhang2021andersson}. For slightly-reduced phases, $15 \leq n \leq 39$, the shear planes are perpendicular to $(132)$ \cite{bursill1969new,bursill1971aggregation,chu1970new,yagi1977312}, with the displacement vector of the shear between adjacent rutile slabs being $1/2[0\bar{1}1]$ \cite{bursill1970displacement,yagi1996electronic}. In these phases, the two-dimensional character of the conducting planes containing Ti $3d$ electrons has the potential to host novel electronic properties \cite{taguchi2010anomalous,lechermann2017oxygen,watanabe2009raman,okamoto2004electronic,szot2011tio2}.

Our observations of needle-like structures forming along the (132) and equivalent crystallographic directions in both in-situ and quenched samples indicate the formation of the $(132)$ Magn\'eli phase. This structure is notoriously difficult to synthesize via conventional annealing, leaving many of its properties unexplored until now \cite{hasiguti1994electrical,hasiguti1970electrical,yagi1996electronic}. Our research bridges this knowledge gap, showing consistent formation of the (132) phase in both S1 and S2 samples despite varied preparation conditions and across a wide range of temperatures and current densities. In particular, the (132) scattering emerges as a dominant feature in the S2 sample, though subtler needle-like patterns along (225) also warrant mention. The relationship of this (225) feature to the current direction or vacuum environment in S2 remains an open question, which our ongoing research seeks to clarify. Additionally, we acknowledge the prior identification of disordered (101), (100), and (725) features in slightly reduced rutile \cite{bursill1969new,hasiguti1994electrical}, highlighting that while other defects do form, the (121) and (132) phases are the most pronounced. 

Besides studies using conventional reduction, our observation of oxygen defects due to electrical current is consistent with earlier work on flash sintered TiO$_2$ \cite{yoon2018measurement,li2019nanoscale,masuda2024optical,charalambous2018situ}. In particular, measurements of elastic properties of in-situ flashing rutile TiO$_2$ revealed anomalous softening which was attributed to the formation of oxygen defects. Under DC current, planar defects were observed \cite{li2019nanoscale} while under AC current, Magn\'eli phases were notably absent \cite{masuda2024optical} which emphasizes the fact that current has substantial impact on the nature of defect formation. Studies using in-situ x-ray diffraction found anomalous displacement of oxygen atoms \cite{yoon2018measurement} and peak splitting \cite{charalambous2018situ}. Large oxygen displacements are consistent with the formation of oxygen defects, while peak splitting is consistent with the formation of an enlarged unitcell containing oxygen defects (c.f. Magn\'eli phases, which have enlarged rutile-like unitcells).

Previous efforts to synthesize Magn\'eli phases through annealing required extensive periods, tens of hours or more at above 1000$^\circ$C, to achieve homogeneously reduced samples \cite{hasiguti1994electrical,hasiguti1970electrical,yagi1996electronic}. In contrast, our research demonstrates that the $(132)$ phase can be rapidly produced in bulk crystals in just minutes at significantly reduced temperatures, approximately 400°C, using flash sintering. This result raises intriguing questions about whether the $(121)$ phase could also be synthesized in single crystals exhibiting greater oxygen deficiency through flash sintering. The successful creation of this phase under such conditions has been documented in ceramic TiO$_2$ \cite{xue2023situ}. 

Comparing our findings with earlier research on slightly reduced rutile, which featured lower oxygen vacancy concentrations ($0.00001 < x < 0.0001$) \cite{yagi1996electronic,hasiguti1994electrical}, provides valuable insights. In these cases, shear planes were either absent or scarcely formed, with electrical transport primarily occurring through hopping between isolated Ti interstitials dispersed throughout the material. The emergence of a few shear planes at higher vacancy concentrations was found to impede hopping, leading to reduced conductivity. However, the pronounced scattering intensity related to the (132) shear planes in our neutron diffuse scattering data indicates a substantially higher $x$ in our samples, signifying a greater concentration of oxygen vacancies. Previous studies estimated activation energies to be around $\sim1$ meV and reported conductivity an order of magnitude higher than that observed in our samples, which possess notably larger oxygen vacancy concentrations. This discrepancy likely stems from the distinct nature of transport mechanisms: in earlier studies, shear planes primarily served to scatter carriers, whereas in our samples, transport appears to be predominantly confined within the planes themselves, akin to behavior seen in more significantly reduced Magnéli phases, such as Ti$_4$O$_7$ \cite{szot2011tio2}.

During flash, it is possible that only existing oxygen vacancies order due to the electric field, but more likely, new vacancies form with excess oxygen diffusing out of the sample and vacancies diffusing into the bulk \cite{bursill1972crystallographic,setvin2015aggregation}. The formation of bulk oxygen defects in rutile is already well known \cite{bursill1969new,bursill1971aggregation,cho2006first}, but it was only recently shown that the formation of oxygen Frenkel defects, a precursor to isolated oxygen vacancies, is enhanced by electric fields \cite{abdelouahed2015relevance}. Under applied electric field, thermal displacements could promote the formation of oxygen defects, leading to enhanced conductivity (due to e.g. impurity levels) and the formation of currents and Joule heating, leading to the formation of even more defects and so on. It is also possible that the rapid heating experienced at the onset of flash, rather than the electric field itself, is responsible for the defects formation \cite{kermani2021ultrafast}, but this avenue of investigation is beyond the scope of this work. 

Before concluding, we mention that the formation of \emph{bulk} defects due to electrical current might be unique to rutile TiO$_2$, which is already known to form bulk oxygen defects (c.f. Ti$_4$O$_7$ Magn\'eli phases), even without electrical currents. In other oxides, defects might remain localized near the surface; this is the subject of ongoing investigation.

We demonstrated that TiO$_2$ can transition from a high-resistance to a low-resistance state through the application of electric current (fig.  \ref{fig:main_fig}b), a change attributed to the formation of oxygen defect planes. The changes induced by current are persistent, but can be reversed by annealing in an oxygen rich environment. The concentration of defects can be controlled via the in-situ current density or by quenching in liquid nitrogen to freeze in defects. The defects are consistent with the formation of Magn\'eli phases in slightly reduced rutile; however, the temperature dependence of conductivity in flashed rutile TiO$_2$ shows several electronic phase transitions that are different from what has been previously observed. The anomalous electronic behaviour suggests that novel electronic physics can be induced by passing electric current through rutile TiO$_2$. There may be promising applications of the electric current-induced switchability of TiO$_2$ for memristor technology.

\paragraph{Acknowledgements}

The authors would like to thank Bekki Mills and Elijah Stevens for help with the design and implementation of the new in-situ sample environment at the Spallation Neutron Source at Oak Ridge National Laboratory.  T.S. and D.R. acknowledge support by the U.S. Department of Energy, Office of Basic Energy Sciences, Office of Science, under Contract No. DE-SC0024117. S.J. and R.R. acknowledge support by the Office of Naval Research under grant No. N00014-18-1-2270. Research at ORNL’s SNS was sponsored by the Scientific User Facilities Division, Office of Basic Energy Sciences, U.S. Department of Energy. T.S. was supported by the U.S. Department of Energy, Office of Science, Office of Workforce Development for Teachers and Scientists, Office of Science Graduate Student Research (SCGSR) program. The SCGSR program is administered by the Oak Ridge Institute for Science and Education for the DOE under contract number DE‐SC0014664. A.P., Y. Z., and G.C. acknowledge support by the National Science Foundation via Grant No. DMR 2204811

\paragraph{Author contributions}

D.R. and T.S. wrote the paper with crucial input from F.Y. T.S. and F.Y. performed the neutron scattering experiments. T.S. and S.J. prepared samples. T.S. analyzed the experimental data. A.P. and Y.Z. did materials properties measurments. G.C., R.R., and D.R. provided guidance. D.R., R.R., and F.Y. conceived the project. This work was supervised by D.R.  

\paragraph{Data Availability}

The datasets generated during this work are available from the authors on reasonable request.

\bibliographystyle{elsarticle-num}

\end{document}